\documentclass[final,5p,times,twocolumn]{elsarticle}

\usepackage{graphicx,color}
\usepackage{amsmath,amssymb,bm}

\def\XXint#1#2#3{{\setbox0=\hbox{$#1{#2#3}{\int}$}
     \vcenter{\hbox{$#2#3$}}\kern-.5\wd0}}

\def\tr{{\rm Tr}}
\def\1{\'{\i}}

\journal{Physics Letters B}

\begin{document}

\begin{frontmatter}
\title{The infrared limit of the SRG evolution and Levinson's theorem}

\author{E. Ruiz Arriola}
\ead{earriola@ugr.es}
\address{Departamento
  de F\'isica At\'omica, Molecular y Nuclear and Instituto Carlos I de
  Fisica Te\'orica y Computacional \\ Universidad de Granada, E-18071
  Granada, Spain}

\author{S. Szpigel}
\ead{szpigel@mackenzie.br}
\address{Centro de R\'adio-Astronomia e Astrof\'\i sica Mackenzie, Escola de Engenharia \\
Universidade Presbiteriana Mackenzie, Brazil}

\author{V. S. Tim\'oteo}
\ead{varese@ft.unicamp.br}
\address{Grupo de \'Optica e Modelagem Num\'erica - GOMNI, Faculdade de Tecnologia - FT \\
Universidade Estadual de Campinas - UNICAMP, Brazil}

\begin{abstract}
On a finite momentum grid with $N$ integration points $p_n$ and
weights $w_n$ ($n=1, \dots N$) the Similarity Renormalization Group
(SRG) with a given generator $G$ unitarily evolves an initial
interaction with a cutoff $\lambda$ on energy differences, steadily
driving the starting Hamiltonian in momentum space $H_{n,m}^{0}= p_n^2
\delta_{n,m}+ V_{n,m}$ to a diagonal form in the infrared limit
($\lambda \to 0$), $H_{n,m}^{G,\lambda \to 0} = E_{\pi(n)}
\delta_{n,m}$, where $\pi(n)$ is a permutation of the eigenvalues
$E_{n}$ which depends on $G$. Levinson's theorem establishes a
relation between phase-shifts $\delta(p_n)$ and the number of
bound-states, $n_B$, and reads $\delta (p_1) - \delta (p_N) = n_B
\pi$. We show that unitarily equivalent Hamiltonians on the grid
generate reaction matrices which are compatible with Levinson's
theorem but are phase-inequivalent along the SRG trajectory.  An
isospectral definition of the phase-shift in terms of an energy-shift
is possible
% as
% $$ \delta (p_n) = - \pi \lim_{\lambda \to 0}
% \frac{H_{n,n}^{{\rm G},\lambda} - p_n^2}{2 w_n p_n} $$
but requires in addition a proper ordering of states on a momentum
grid such as to fulfill Levinson's theorem. We show how the SRG with
different generators $G$ induces different isospectral flows in the
presence of bound-states, leading to distinct orderings in the
infrared limit. While the Wilson generator induces an ascending
ordering incompatible with Levinson's theorem, the Wegner generator
provides a much better ordering, although not the optimal one. We
illustrate the discussion with the nucleon-nucleon ($NN$) interaction
in the $^1S_0$ and $^3S_1$ channels.
\end{abstract}

\end{frontmatter}

\section{Introduction}

During the last decade the renormalization group equations have
advantageously been used as a technique to simplify microscopic large
scale calculations in Nuclear Structure and Reactions. More
specifically, the Similarity Renormalization Group (SRG) has been
intensively applied to handle multinucleon forces in order to soften
the short-distance core~\cite{Bogner:2006pc,Furnstahl:2013oba} with a
rather universal pattern for nuclear symmetries~\cite{Timoteo:2011tt,
  Arriola:2013nja} and interactions~\cite{Dainton:2013axa}. The basic
strategy underlying the SRG method is to evolve a starting (bare)
interaction $H_0$ which has been fitted to nucleon-nucleon ($NN$)
scattering data via a continuous unitary transformation that runs a
cutoff $\lambda$ on energy differences. Such a transformation
generates a family of unitarily equivalent smooth interactions
$H_{\lambda} = U_{\lambda}H_0U_{\lambda}^{\dagger}$ with a
band-diagonal structure of a prescribed width roughly given by the SRG
cutoff $\lambda$.  For most cases of interest a finite momentum grid
with $N$ integration points $p_n$ and weights $w_n$ ($n=1, \dots N$)
is needed to solve the SRG flow equations numerically, and for such a
finite basis the SRG transformation corresponds to a continuum
generalization of the well-known gauss reduction method of a matrix to
the diagonal form.

Unfortunately the $NN$ force is not yet known from first principles and
most information on the $NN$ interaction is strongly constrained by the
abundant $np$ and $pp$ scattering data~(see e.g. Ref.~\cite{Perez:2013jpa}
for a recent upgrade and references therein). Roughly speaking this is
equivalent to know the phase-shifts with their uncertainties at some center-of-mass ($CM$)
momenta and in a given range, $0 < \Delta p \equiv p_1 < \dots < p_N
\equiv \Lambda $, and in fact a common practice has been to tabulate
the phase-shifts at given discrete set of energy values. The implicit
assumption underlying this practice is that one expects this discrete
information to encode and summarize sufficient details on the
interaction, in full harmony with the need of solving SRG flow equations on
a finite grid. The computational advantages of using properly chosen
few discrete variables for finite volume systems such as
nuclei~\cite{Bulgac:2013mz} have been emphasized as the number of
states gets drastically reduced.

In a previous note~\cite{Arriola:2013gya} we have
suggested to pursue this SRG transformation to the very limit since
this naturally drives the interaction to a diagonal form and hence
removing all off-shell ambiguities. The two most common choices for
the SRG generator which guarantee that the SRG flow equations
evolve the Hamiltonian to the diagonal form are the so-called Wegner
and Wilson generators. The non-trivial question pertains the ordering
of states arising in general from any diagonalization procedure and from
the SRG flow equations in particular. On a finite momentum grid the SRG evolution with both Wilson and Wegner generators can drive the Hamiltonian to the diagonal form when $\lambda\to0$ (unless degeneracies appear in the diagonal). However, in the case of Wegner generator all $N!$ possible permutations corresponding to the final ordering of the eigenvalues are stable fixed points while in the case of Wilson generator only the permutation in which the eigenvalues are in ascending order is a stable fixed point. A perturbative asymptotic fixed point
analysis~\cite{Timoteo:2011tt,Arriola:2013gya} provides an analytical
understanding of the phenomenon observed in the numerical
calculations.

On a finite momentum grid the scattering process becomes a bound-state
problem~\cite{muga1989stationary} and many important properties such
as the intertwining properties of the Moller wave operators do not
hold. Actually, we will show that on the momentum grid the reaction matrix
generally used to solve the Lippmann-Schwinger (LS) equation
does not produce isospectral phase-shifts, i.e. $\delta^ {\rm LS} (H_0)
\neq \delta^ {\rm LS} (U_{\lambda}H_0U_{\lambda}^{\dagger})$. There has been a renewed and
continuous effort by Kukulin {\it et al.}~\cite{kukulin2009discrete,Rubtsova:2010zz} to formulate a new
approach toward a direct evaluation of the multichannel multienergy $S$-matrix without solving the scattering equations in the few-body problem. These attempts can be traced from early work by Lifshits in
the 1940's (see e.g.~\cite{kukulin2009discrete,Rubtsova:2010zz}) where
quite generally the relevance of the energy-shift was established for
impurities in a crystal. This is similar to the relation between the
energy-shift and the phase-shift at large
volumes~\cite{Fukuda:1956zz,DeWitt:1956be}. Because of more recent
popularity within lattice QCD calculations it is called the Luscher
formula when the momentum grid is fixed by the finite lattice
volume~\cite{Luscher:1985dn,Luscher:1990ux}. The energy-shift approach
does comply to the isospectrality requirement, as it just involves the
eigenvalues. In the continuum limit all these approaches are expected
to fulfill Levinson's theorem~(see e.g. \cite{Ma:2006zzc} and
references therein).

In the present paper we want to display an interesting connection
between the SRG method in the infrared limit~\cite{Arriola:2013gya},
which drives the system to a diagonal interaction, and the eigenvalue
method for scattering which can be formulated without any reference to
the SRG flow and the scattering equations. For a finite dimensional
space with dimension $N$, there are $N!$ possible orderings for the eigenvalues of the diagonalized Hamiltonian and, as we will show, when bound-states occur picking the
right ordering proves crucial to establish an energy-shift which
allows to deduce phase-shifts embodying Levinson's
theorem~\cite{Ma:2006zzc} in the continuum limit.

\section{SRG on a momentum grid}

The general SRG flow equation corresponds to a one-parameter operator
evolution dynamics given by~\cite{Kehrein:2006ti},
\begin{eqnarray}
\frac{d H_s}{ds} = [[ G_s, H_s],H_s] \; ,
\label{eq:SRG}
\end{eqnarray}
and supplemented with a boundary condition, $\lim_{s \to 0 }H_s = H_0$. As it is
customary we will often switch from the flow parameter $s$ to the SRG cutoff variable
$\lambda=s^{-1/4}$ which has dimensions of momentum. The isospectrality of
the SRG becomes evident from the trace invariance property $\tr
(H_s)^n= \tr (H_0)^n$. The SRG generator $G_s$ can be chosen according to
certain requirements, and here we will use two popular choices: the
relative kinetic energy $G_s=T$, which is by construction independent of s
~\cite{wegner1994flow} (Wilson generator), and
the evolving diagonal part of the Hamiltonian $G_s=diag(H_s)$
~\cite{Glazek:1994qc} (Wegner generator). %or a block-diagonal (BD) generator $H^{BD}_s=PH_sP+QH_sQ$ where $P+Q=1 $ are orthogonal projectors $P^2 = P$, $Q^2 = Q$ , $QP=PQ=0$, for states below and above a given momentum~\cite{Anderson:2008mu}%.
Normalizations are taken as in Refs.~\cite{Szpigel:2010bj}.

For simplicity we consider the toy model separable gaussian potential
discussed previously~\cite{Arriola:2013era,Arriola:2013yca} which provides a
reasonable description of the $NN$ system in the $^1S_0$ and $^3S_1$
partial-wave channels at low-momenta.  The action of the (bare)
Hamiltonian on a given state in momentum space is given by (here and
in what follows we use units such that $\hbar=c=M=1$, where $M$ is the
nucleon mass)
\begin{eqnarray}
H_0 \psi (p)=
p^2 \psi(p) + \frac{2}{\pi} \int_0^\infty q^2 dq V_0 (p,q) \psi(q) \; .
\label{eq:schro}
\end{eqnarray}

The SRG flow equations are solved numerically on a $N$-dimensional
momentum grid, $p_1 < \dots < p_N $, by implementing a high-momentum
ultraviolet (UV) cutoff, $p_{\rm max}=\Lambda$, and an infrared (IR)
momentum cutoff $p_{\rm min} = \Delta p$ ~\cite{Szpigel:2010bj}. The
integration rule becomes
\begin{eqnarray}
\int_{\Delta p}^\Lambda dp f(p) \to \sum_{n=1}^N w_n f(p_n) \, .
\end{eqnarray}
The SRG flow equations on the grid follow from inserting the completeness
relation in discretized momentum space
\begin{eqnarray}
1=\frac{2}{\pi}\sum_{n=1}^N w_n p_n^2 | p_n \rangle \langle p_n | \, .
\end{eqnarray}
For instance, the eigenvalue problem on the grid may be formulated as
\begin{eqnarray}
H_{\lambda} \varphi_\alpha (p) = P_\alpha^2 \varphi_\alpha (p)   \, ,
\end{eqnarray}
where the matrix representation of the Hamiltonian reads
\begin{eqnarray}
H_{\lambda}(p_n,p_m) = p_n^2 \delta_{n,m} + \frac{2}{\pi} w_n p_n^2 V_{\lambda}(p_n,p_m)  \, .
\end{eqnarray}
A bound-state with (negative) eigenvalue $P_\alpha^2 =- B_\alpha$ corresponds to a pole in the scattering amplitude at imaginary momentum $P_\alpha= i \gamma$.

Because of the commutator structure of the SRG flow equation the isospectrality property still holds
on the grid, i.e. $H_{\lambda} = U_{\lambda}H_0U_{\lambda}^{\dagger}$, and therefore the eigenvalues $P^{2}_{\alpha}$ of $H_{\lambda}$ are $\lambda$-independent,
\begin{eqnarray}
\frac{d P_\alpha}{d \lambda}=0  \, .
\end{eqnarray}

Although the eigenvalues are preserved along the SRG trajectory, the ordering of the states depends on the generator $G$ of the SRG transformation. A lot of accumulated numerical experience has shown that in the presence of bound-states (real or spurious) Hamiltonians evolved using Wilson and Wegner generators start behaving differently when the SRG cutoff $\lambda$ approaches some critical momentum $\Lambda_c$, which corresponds to the threshold scale where the bound-state emerges \cite{Glazek:2008pg,Wendt:2011qj}. In the Wilson generator case the bound-state remains coupled to the low-momentum scales as $\lambda$ approaches $\Lambda_c$, such that the bound-state eigenvalue is pushed towards the lowest momentum available on the grid, which corresponds to the IR momentum cutoff $\Delta p$. Moreover, when $\Delta p \to 0$  matrix-elements of the potential at low-momentum diverge in order to force the bound-state eigenvalue to smaller momenta, such that the SRG evolution may become numerically unstable. In the Wegner generator case the bound-state decouples from the low-momentum scales as $\lambda$ approaches $\Lambda_c$, being placed on the diagonal of the Hamiltonian as an isolated negative eigenvalue at a momentum between the lowest momentum on the grid and $\Lambda_c$. As pointed out in Ref.~\cite{Wendt:2011qj}, the {\it a priori} determination of the position at which the bound-state is placed on the diagonal when using Wegner generator is still an open problem. It is important to note that when the SRG cutoff $\lambda$ is kept well above $\Lambda_c$ or in the absence of bound-states the SRG evolution using Wilson and Wegner generators are nearly identical, a behavior that can be traced to the dominance of the kinetic energy.
\begin{figure}[t]
\begin{center}
\includegraphics[width=8.5cm]{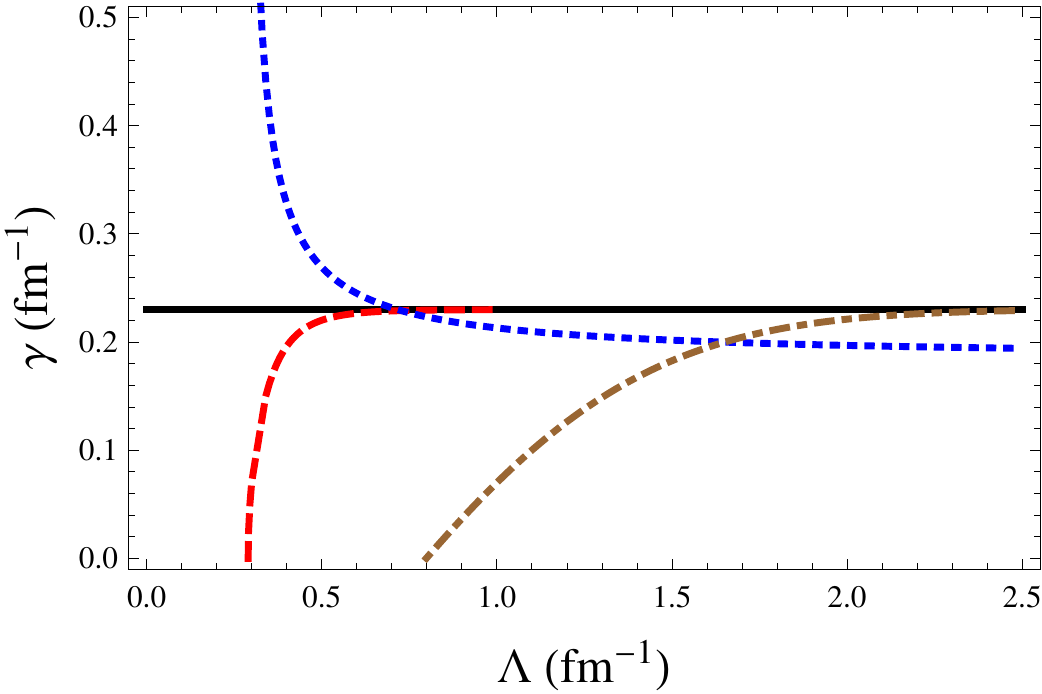}\hspace{0.5cm}
\end{center}
\caption{Variational (dotted-dashed) and implicit (LO: dotted, NLO: dashed) determinations of the critical
  momentum scale $\Lambda_c$ above which the interaction supports a bound state,
  $P^2 = - B = - \gamma (\Lambda)^2$.  The full Hamiltonian has
  $\lim_{\Lambda \to \infty} \gamma (\Lambda)= \gamma_d =0.23~ {\rm
    fm}^{-1}$ corresponding to the deuteron (solid). The critical value $\Lambda_c$
  corresponds to $\gamma=0$. Note that in the LO case, since the interaction is purely
  attractive, the threshold cutoff gives $\gamma = \infty$.}
\label{fig:lambd-crit}
\end{figure}

%implicit analysis

One should note that the critical momentum scale $\Lambda_c$ is
distinct from the characteristic bound-state momentum scale $\gamma$.
For weakly coupled bound-states, such as the deuteron, we can make an
estimation of $\Lambda_c$ by exploiting the complementarity between
the implicit and explicit renormalization of effective interactions
analyzed in Ref.~\cite{Arriola:2013era}. This is based on using
low-energy scattering data to encode the high-energy part of the
interaction by imposing suitable renormalization conditions for an
effective theory with a momentum cutoff scale $\Lambda$ that divides
the Hilbert space into a low-momentum $P$-space ($p < \Lambda$) and a
high-momentum $Q$-space ($p > \Lambda$), separating explicitly what
degrees of freedom and interactions behave dynamically. 
At low values
of $\Lambda$ the interaction can be expanded in powers of momenta
($p,p' < \Lambda$),
\begin{eqnarray}
V(p',p) = C_0 (\Lambda) + C_2(\Lambda) (p^2+p'^2) + \dots \; .
\end{eqnarray}

We can determine the low-energy constants $C_0, C_2, \dots$ from low-energy data. For instance at leading-order (LO) we just fix the scattering length $\alpha_0$ at {\it any} value of $\Lambda$ which
leads to the running of the coupling constant $C_0$ given by
\begin{eqnarray}
C_0(\Lambda)= \frac{\alpha_0}{1-\frac{2\Lambda \alpha_0}{\pi}} \ .
\end{eqnarray}

In this simple contact theory the deuteron wave function is given by
the equation
\begin{eqnarray}
\Psi_d(p) = \frac{Z}{p^2+\gamma^2} \; ,
\label{eq:deut-wf}
\end{eqnarray}
where $Z$ is determined from the normalization condition of the bound-state and fulfills the relation
\begin{eqnarray}
Z \left(1+ \frac{2}{\pi} C_0(\Lambda) \int_0^\Lambda \frac{q^2}{q^2+\gamma^2} dq
\right) =0 \; .
\end{eqnarray}
Clearly, in order to get a non trivial solution $Z\neq 0$ the coupling
constant $C_0$ must be negative. As we see this requires $\alpha_0 >
0$ and $\Lambda > \Lambda^{\rm LO}_c = 2/\pi \alpha_0$. Taking
$\alpha_0=5.42~{\rm fm}$ for the $^3S_1$ channel we obtain
$\Lambda^{\rm LO}_c\sim 0.3 ~{\rm fm}^{-1}$. The calculation at
next-to-leading-order (NLO) further determines $C_2(\Lambda)$ from the
effective range $r_0=1.75~{\rm fm}$~\cite{Arriola:2013era} and the
deuteron wave function in Eq.~(\ref{eq:deut-wf}) is modified by
replacing $Z\to Z(1+ X q^2)$ and after solving for $X$ provides a tiny
correction, $\Lambda^{\rm NLO}_c = 0.29~ {\rm fm}^{-1}$.  A different
variational estimate can be done by looking at what $\Lambda$ the
matrix Hamiltonian in the P-space  supports a bound state.

The emergence of the threshold scale $\Lambda_c$ is displayed in
Fig.~(\ref{fig:lambd-crit}), where the characteristic deuteron
momentum scale $\gamma$ is shown as a function of the cutoff $\Lambda$
for the variational and the implicit renormalization estimates. In
Fig.~(\ref{fig:lambd-crit}) we also see that the NLO approximation
saturates at the exact value of $\gamma$ for $ \Lambda \sim 2
\Lambda_c$. The performance of the variational approach, for the model
under study, is not good. Actually, large values of $\Lambda$ are
needed to saturate the bound state. Note that the exact solution would
have a discontinuity at the exact $\Lambda_c$.

% 3S1 H crossing

%
\begin{figure*}[t]
\begin{center}
\includegraphics[width=8.5cm]{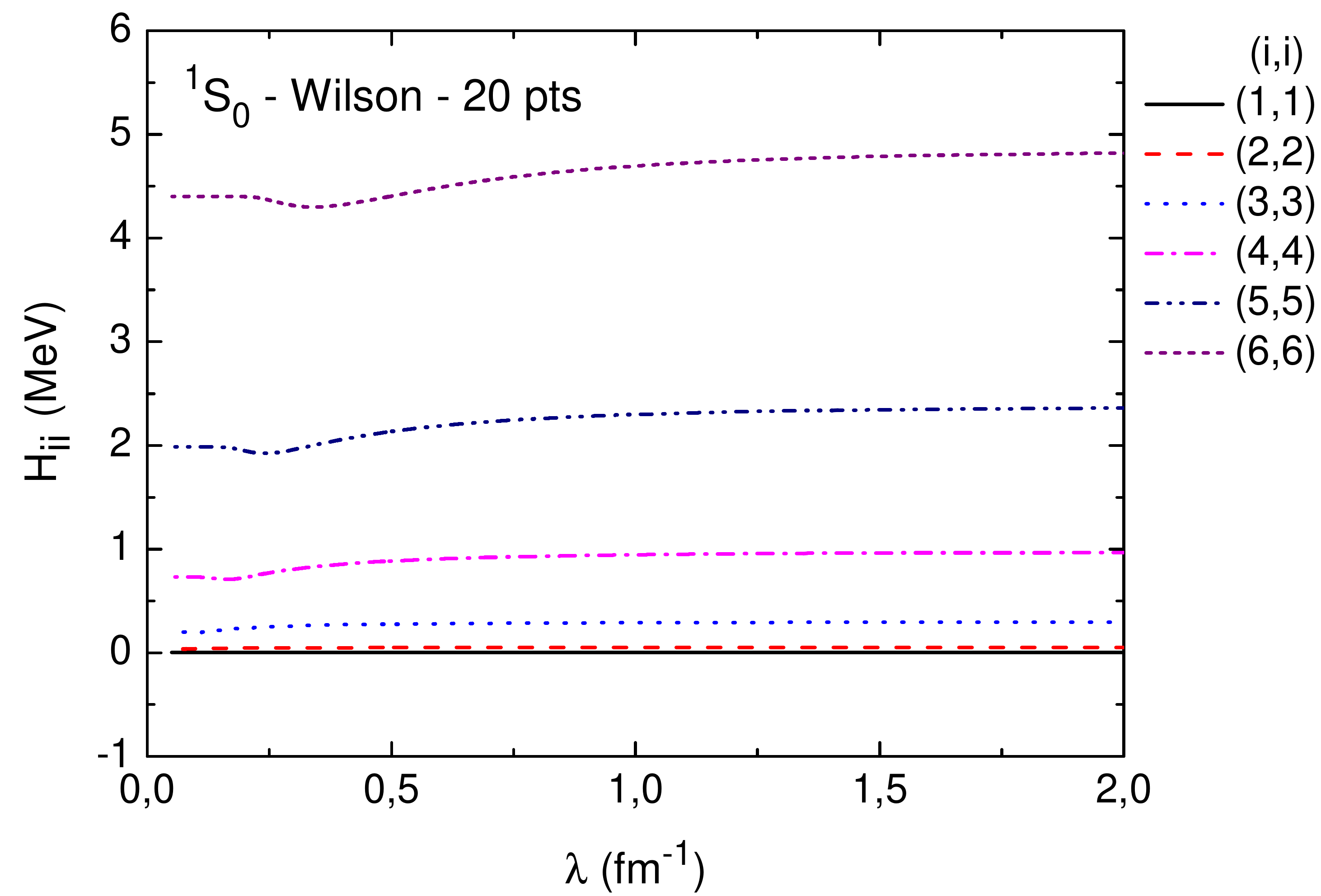}\hspace{0.5cm}
\includegraphics[width=8.5cm]{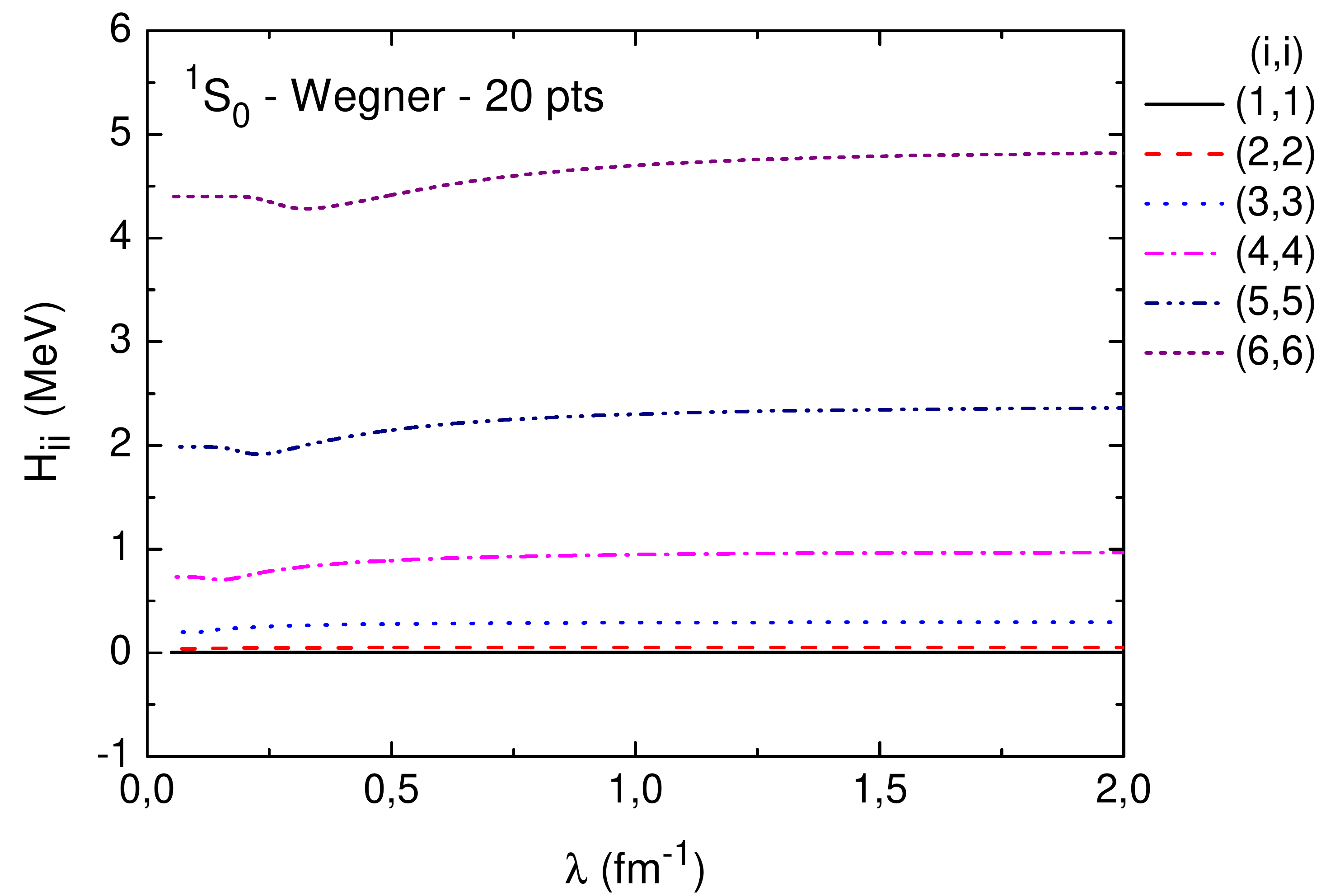} \\ \vspace{0.5cm}
\includegraphics[width=8.5cm]{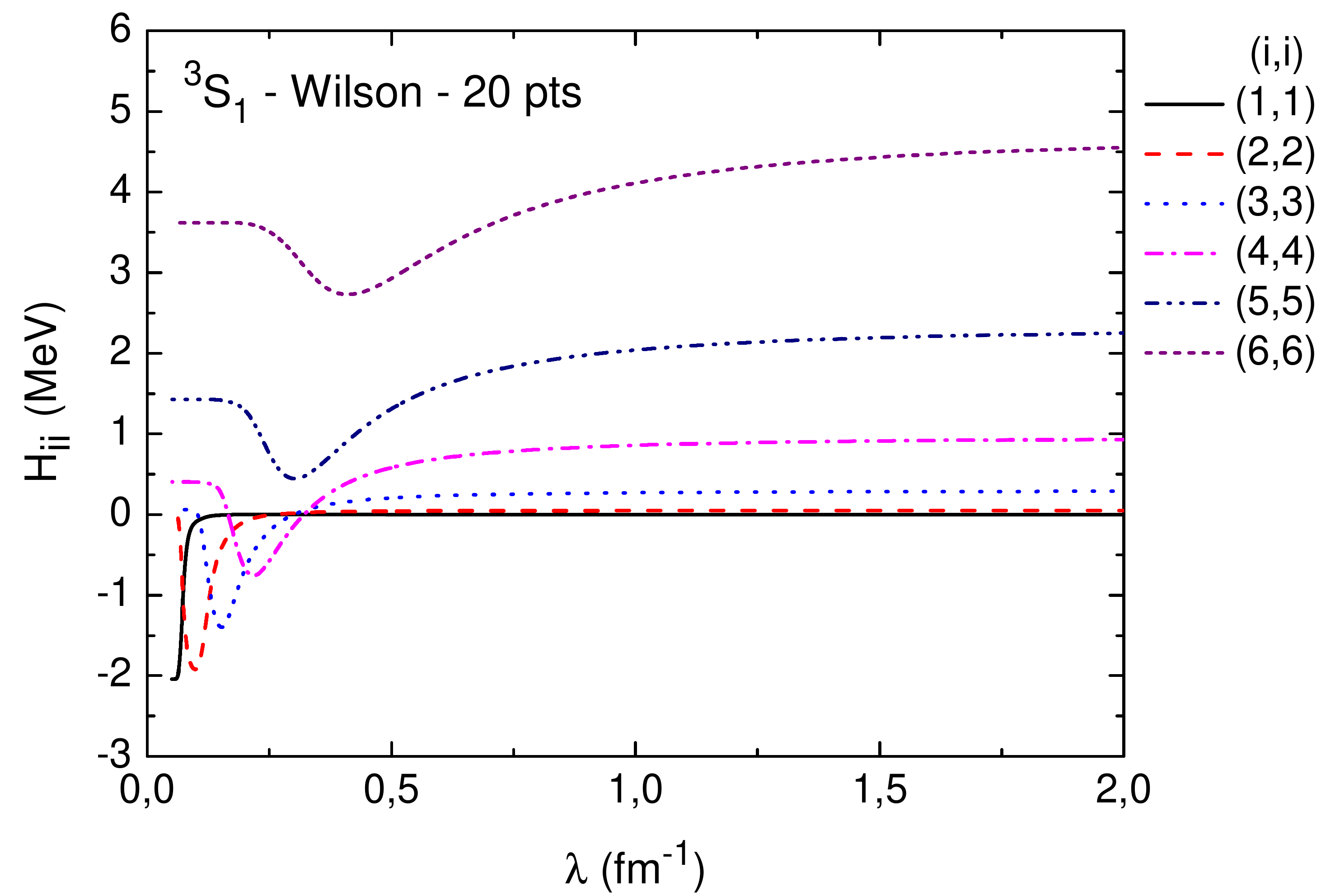}\hspace{0.5cm}
\includegraphics[width=8.5cm]{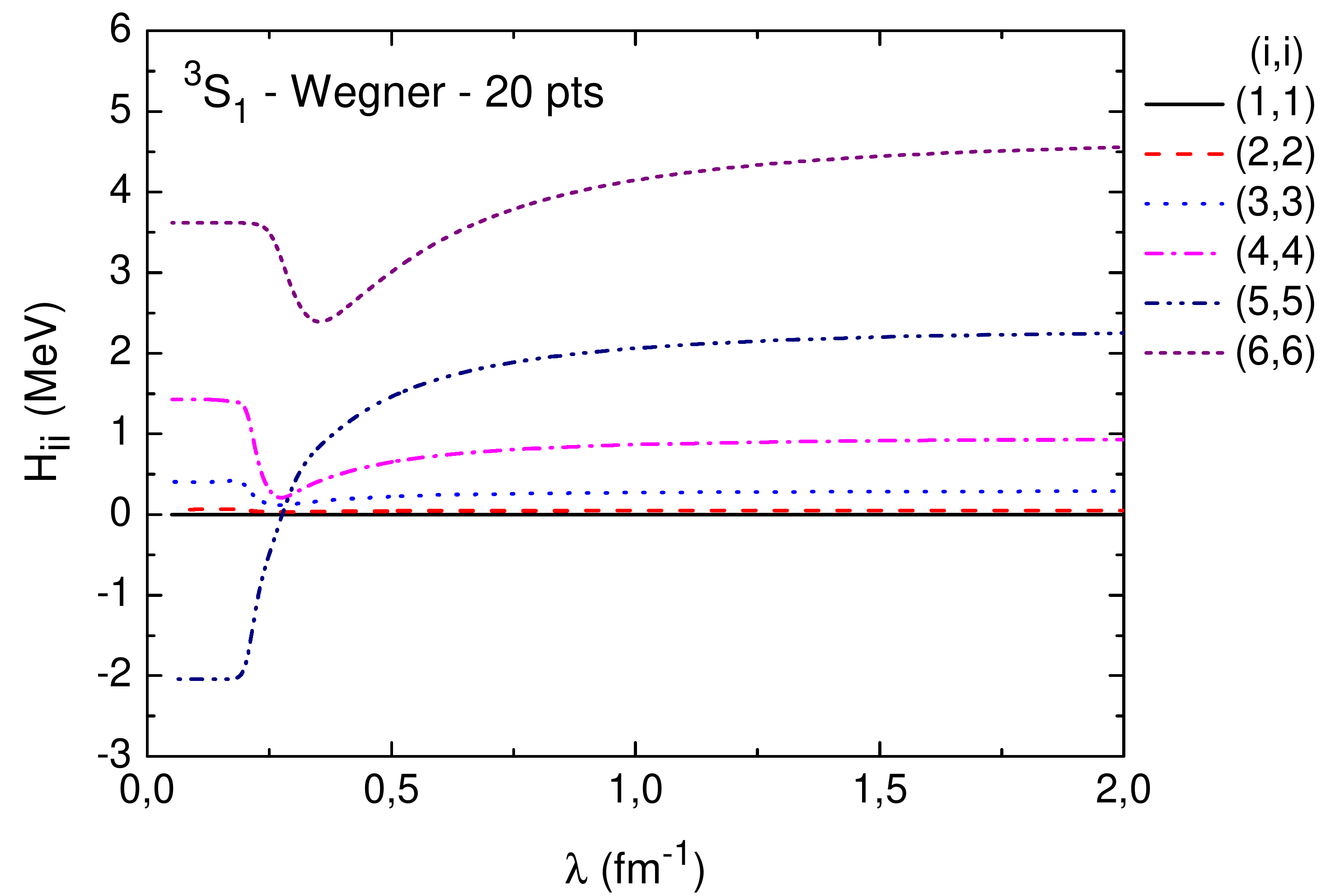}
\end{center}
\caption{SRG evolution of the diagonal matrix-elements of the
  hamiltonian for the toy model potential in the $^1S_0$ (upper panels)
and $^3S_1$ (lower panels) channels using
  the Wilson (left) and the Wegner (right) generator. We have considered a high-momentum UV
  cutoff $\Lambda=2~{\rm fm}^{-1}$ and $N=20$ grid points. The SRG cutoff $\lambda$ was varied in a range from $0.05$ to $2.0 {\rm fm}^{-1}$}
\label{fig:25}
\end{figure*}

In Fig.~(\ref{fig:25}) we show the SRG evolution of the lowest
diagonal matrix elements of the toy model hamiltonian
$H_{n,n}^{G,\lambda}$ (n=1,...,6) in the $^1S_0$ and $^3S_1$
partial-wave channels, for a momentum grid with $N=20$ points and
$\Lambda=2~{\rm fm}^{-1}$. The SRG cutoff $\lambda$ was varied in a
range from $0.05$ to $2.0~{\rm fm}^{-1}$. An interesting difference
can be observed between the SRG evolution with Wilson and Wegner
generators in the infrared limit. As the SRG cutoff $\lambda$
approaches the critical momentum scale $\Lambda_c$, there is no
crossing amongst the diagonal matrix elements for the $^1S_0$ channel
both with Wilson and Wegner generators indicating that the initial
ascending order is maintained all along the SRG trajectory. Moreover,
both generators lead to similar SRG evolutions, as expected since
there are no bound-states. For the $^3S_1$ channel, on the other hand,
there are crossings with both generators. In the Wilson generator case
the initial ascending order is asymptotically restored in the limit
$\lambda\to 0$ ~\cite{Timoteo:2011tt,Arriola:2013gya} with the lowest
momentum diagonal matrix element $H_{\lambda}(p_1,p_1)$ flowing into
the deuteron bound-state. In the Wegner generator case a re-ordering
occurs with some upper momentum diagonal matrix element
$H_{\lambda}(p_{n_{\rm {BS}}},p_{n_{\rm {BS}}})$ flowing into the
deuteron bound-state in the limit $\lambda\to 0$. As shown in
Ref.~\cite{Wendt:2011qj} for LO chiral effective field theory (EFT)
interactions with large momentum cutoffs $\Lambda_{\rm EFT}$, the
position at which the (spurious) bound-state is placed changes with
the cutoff. In our calculations for the separable gaussian toy model
potential on a finite momentum grid we observe a similar change of the
bound-state position when using different values for the number of
grid points $N$ and/or the high-momentum UV cutoff $\Lambda$. For the
calculation with $N=20$ grid points and $\Lambda=2~{\rm fm}^{-1}$,
shown in Fig.~(\ref{fig:25}), the momentum at which the bound-state is
placed corresponds to $p_{n_{\rm BS}}\to p_5 \sim 0.254~{\rm
  fm}^{-1}$. As one can observe, the diagonal matrix-element
$H_{\lambda}(p_{BS},p_{BS})$ that flows into the deuteron bound-state
is the one that starts to decrease rapidly towards negative values
when the SRG cutoff $\lambda$ approaches $\Lambda_c \sim 0.3 ~{\rm
  fm}^{-1}$, indicating the break-up of the kinetic energy dominance,
i.e.
\begin{eqnarray}
p_{n_{\rm {BS}}}^2 < \frac{2}{\pi}~w(p_{n_{\rm {BS}}})~p_{n_{\rm {BS}}}^2~|V_{\lambda < \Lambda_c}(p_{n_{\rm {BS}}},p_{n_{\rm {BS}}})|  \; .
\end{eqnarray}

\begin{figure*}[t]
\begin{center}
\includegraphics[width=8cm]{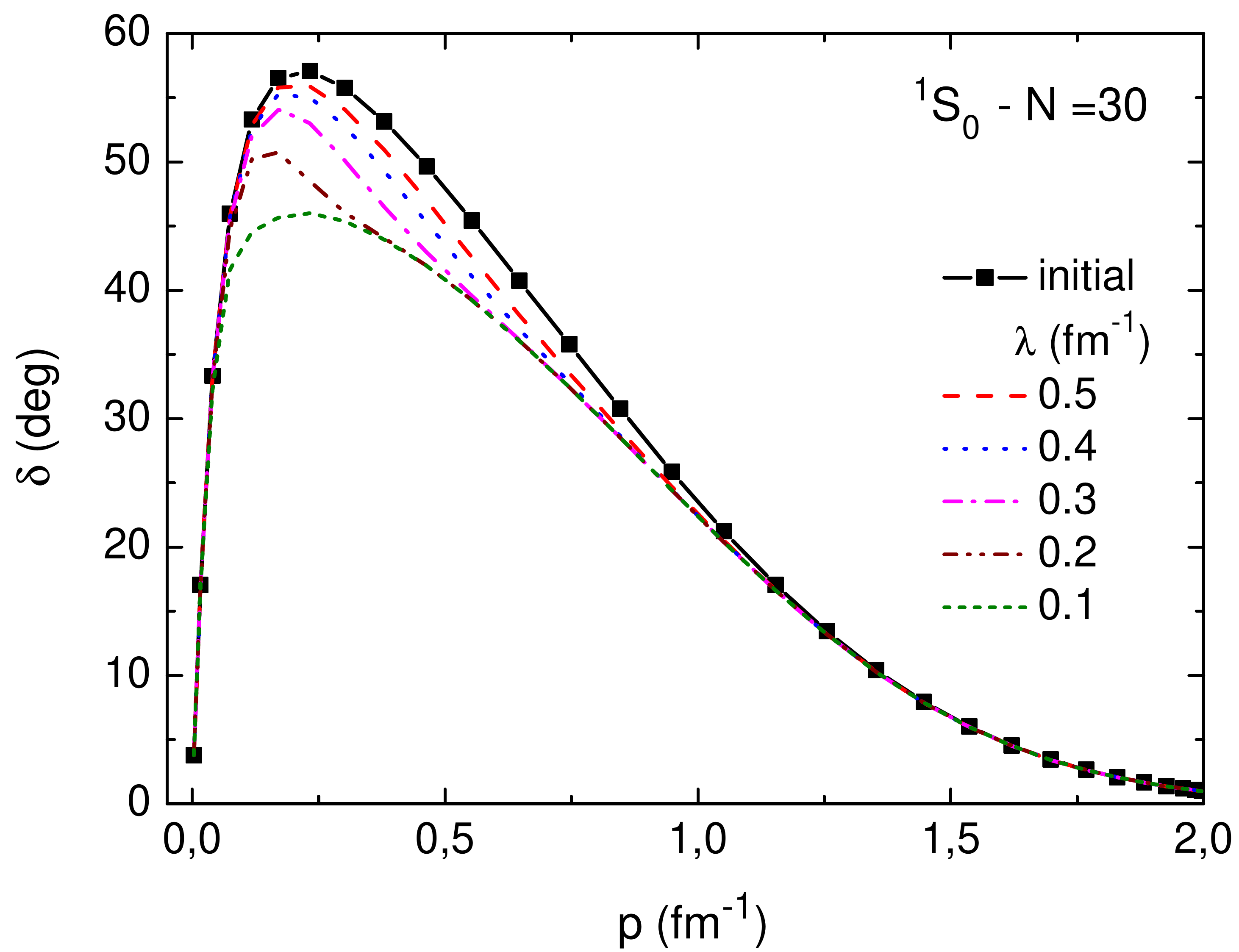}\hspace{0.5cm}
\includegraphics[width=8.2cm]{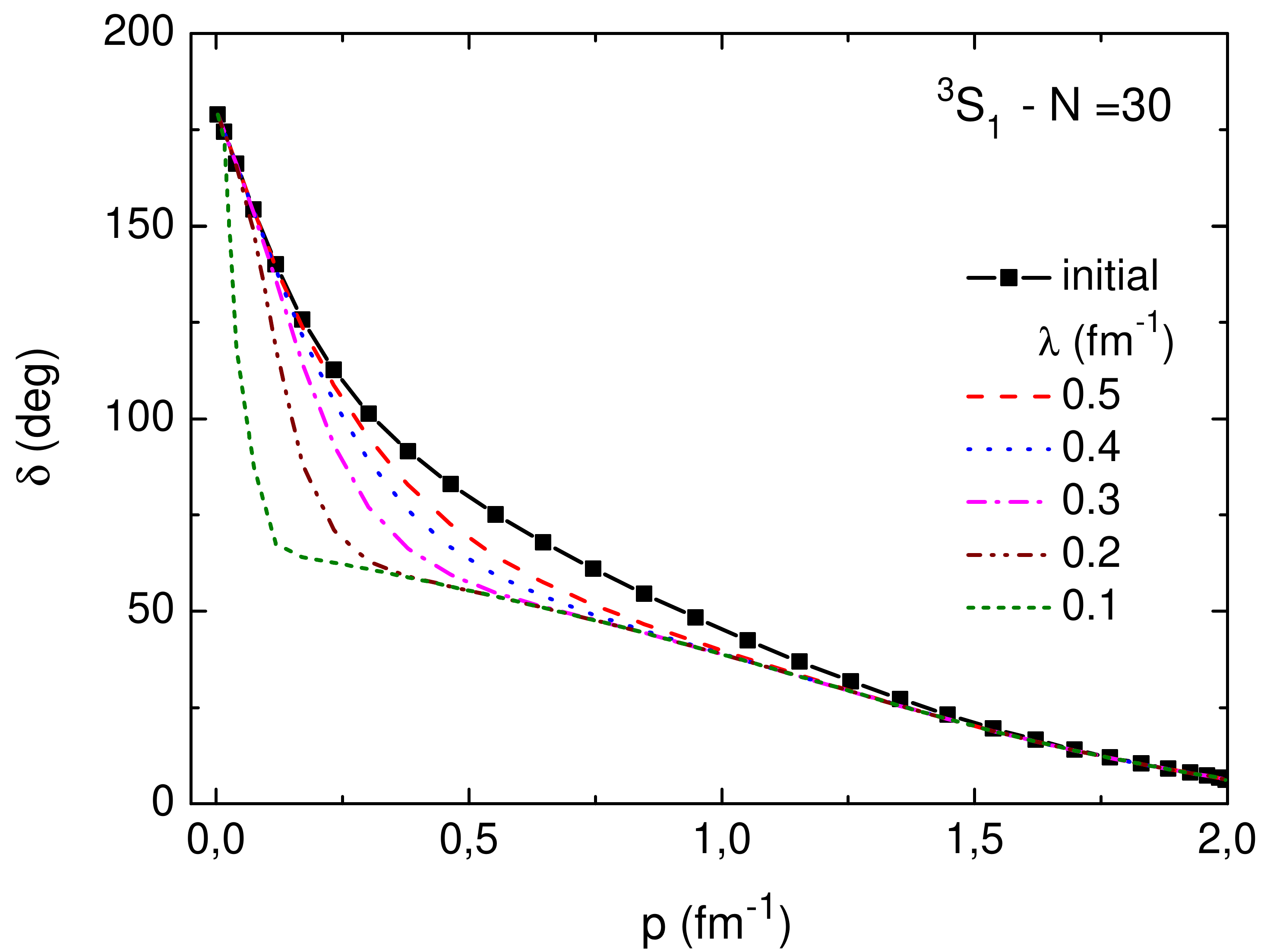}\hspace{0.5cm}
\end{center}
\caption{Phase-shifts evaluated from the solution of the LS equation on the momentum grid with the toy model potential in the $^1S_0$ (left) and $^3S_1$ (right) channels evolved through the SRG transformation with Wilson generator for several values of the SRG cutoff $\lambda$. We have considered a high-momentum UV cutoff $\Lambda=2~{\rm fm}^{-1}$ and $N=30$ grid points.}
\label{fig:ps-noninv}
\end{figure*}

\section{Phase-inequivalence  of the reaction matrix on a momentum grid}

As mentioned above the original motivation for the SRG method was to
soften the interaction while keeping the phase-shifts invariant.  As
we will show below the verification of phase-equivalence along the SRG
trajectory requires a proper definition of the phase-shift in a
momentum grid. This is a subtle point, particularly when the
interaction is attractive enough to generate bound states.

The standard procedure so far within the SRG approach has been to
solve the Lippmann-Schwinger (LS) equation for the $T$-matrix. In operator
form the LS equation reads
\begin{eqnarray}
T= V + V (p^2 - H_0- i \epsilon)^ {-1} T \; .
\end{eqnarray}
Taking matrix elements on the momentum grid we get
\begin{eqnarray}
T_{nm}(p) = V_{nm} + \frac2{\pi} \sum_{k=1}^N w_k \frac{p_k^2}{p^2-p_k^2+ i \epsilon} V_{nk} T_{k,m}(p) \; .
\end{eqnarray}
where $p^2$ is the scattering energy. The on-shell limit is obtained
by taking $p=p_l$ on the grid. As usual we switch to the reaction
matrix which on the grid yields the equation for the half-on-shell amplitude
\begin{eqnarray}
R_{nm}(p_m) = V_{nm} + \frac2{\pi} \sum_{k \neq m} w_k \frac{p_k^2}{p_m^2-p_k^2} V_{nk} R_{k,m}(p_m) \; ,
\end{eqnarray}
where the excluded sum embodies the principal value prescription of
the continuum version. This equation can be solved by inversion for
any grid point $p_n$ and thus we may obtain the phase-shifts
\begin{eqnarray}
-\frac{\tan \delta^{\rm LS}(p_n)}{p_n} = R_{nn} (p_n) \; ,
\label{eq:ps-LS}
\end{eqnarray}
where the supper-script LS denotes that these phase-shifts are obtained from
the solution of the LS equation on the grid. Of course, the limit $N
\to \infty$ should be understood in the end.

Let us analyze the behavior of the phase-shifts as computed from the
definition given in Eq.~(\ref{eq:ps-LS}) using the potentials
$V_{nm}(\lambda)$ evolved according to the SRG flow equations, Eq.~(\ref{eq:SRG}),
on the finite grid. The results for the toy model potential in the $^1S_0$ (left) and $^3S_1$ (right) channels are presented in Fig.~(\ref{fig:ps-noninv}) for a high-momentum UV cutoff $\Lambda=2~{\rm fm}^{-1}$, $N=30$ grid points and several values of the SRG cutoff $\lambda$. As we see, Levinson's theorem~\cite{Ma:2006zzc}, $\delta_\lambda^{\rm LS}(p_1)-\delta_\lambda^{\rm LS}(p_N)= n_B \pi$, is fulfilled on the
grid. However, while this discretization enables to handle SRG flow
equations numerically, the price to pay due to the finite momentum
grid, however, is that on this grid the phase-shifts as obtained from
the LS equation are not independent of the SRG cutoff variable
$\lambda$. While the lack of phase-equivalence disappears for large
$N$ we want to analyze the possibility whether one can define
SRG-independent phase-shifts on the grid for {\it any} value of the
dimension $N$.
\begin{figure*}[t]
\begin{center}
\includegraphics[width=8cm]{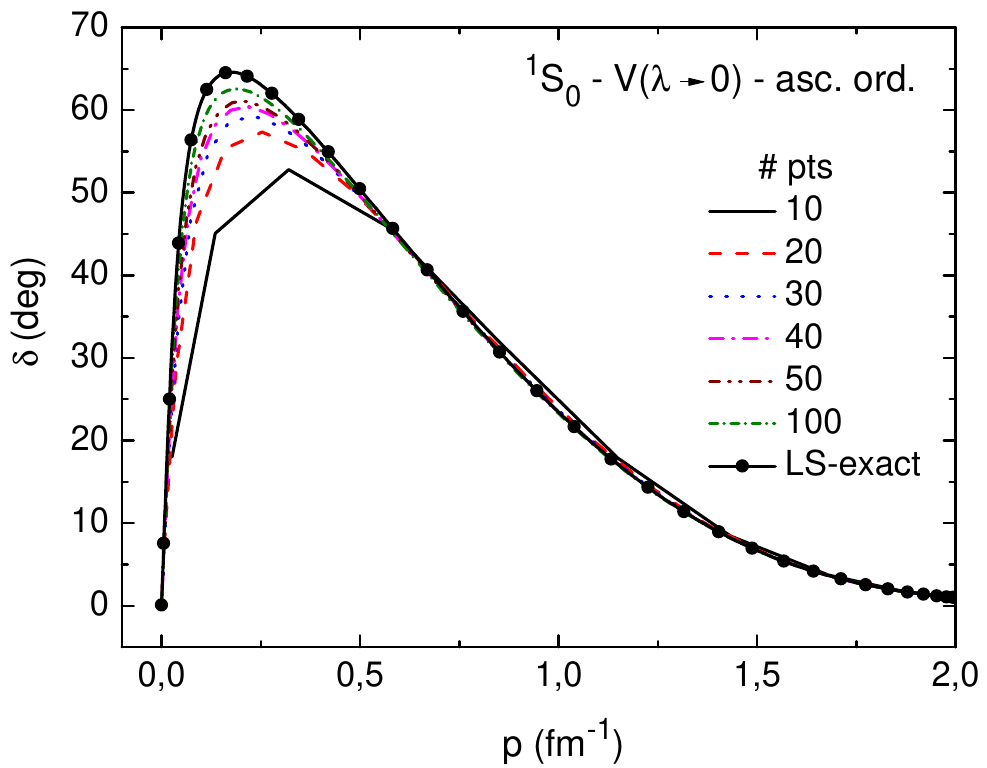}\hspace{0.5cm}
\includegraphics[width=8cm]{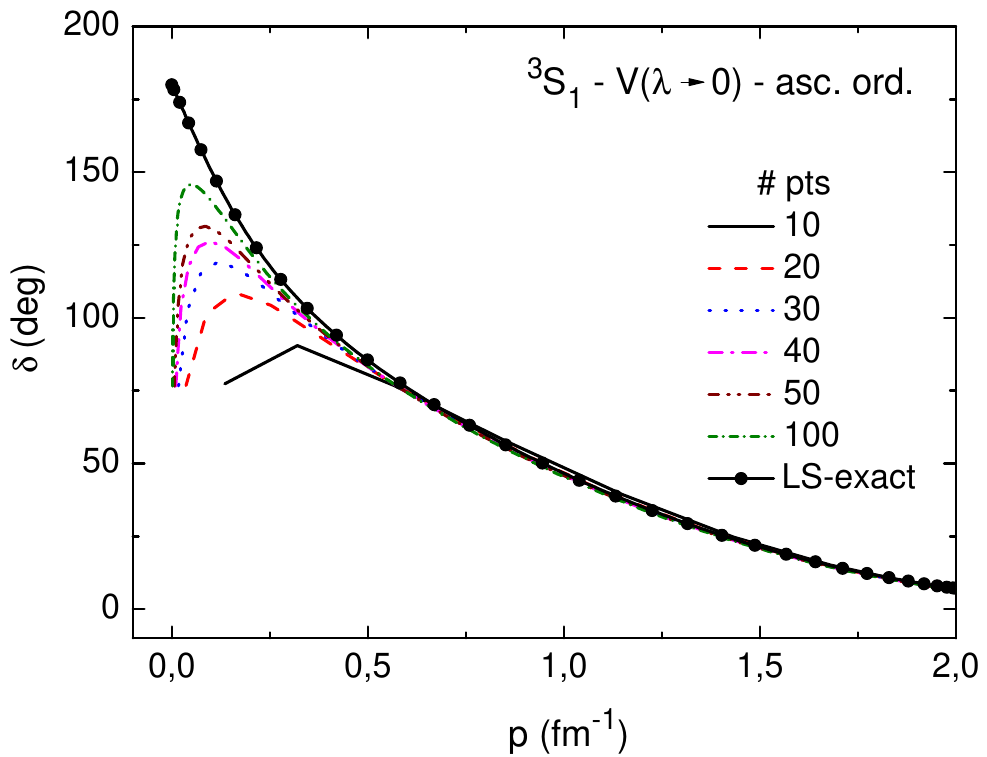}
\\
\includegraphics[width=8cm]{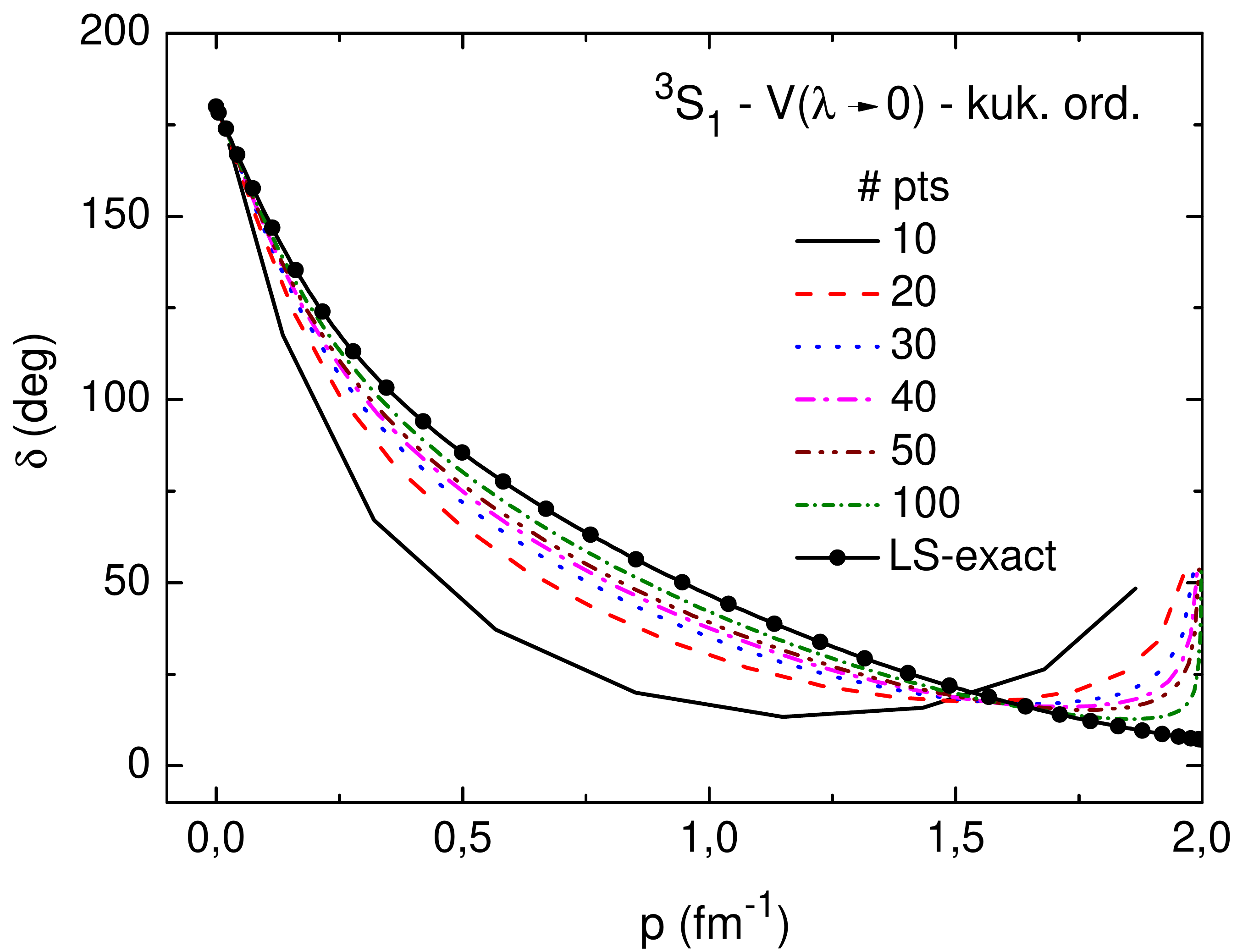}\hspace{0.5cm}
\includegraphics[width=8cm]{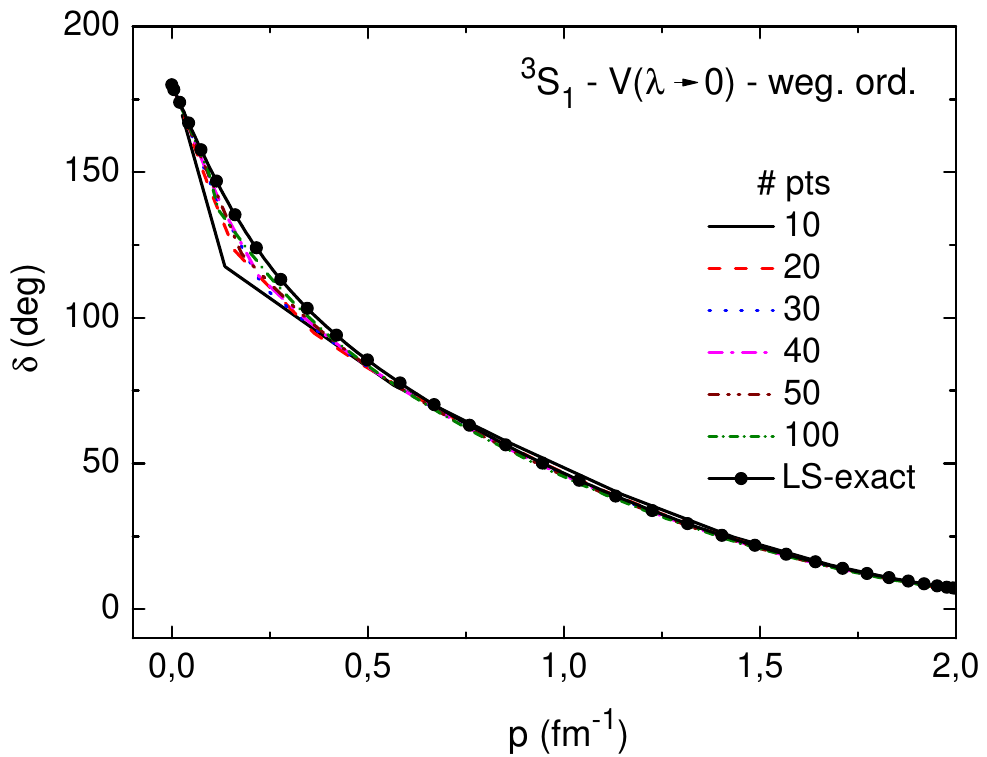}
\end{center}
\caption{Phase-shifts for the toy model potential in the $^1S_0$ and
  $^3S_1$ channels evaluated by the eigenvalue method with the
  eigenvalues sorted in several ways. Upper left panel: $^1S_0$
  channel in ascending order. Upper right panel: $^3S_1$ channel in
  ascending order.  Lower left panel: $^3S_1$ channel with
  Kukulin {\it et al.} order.  Lower right panel: $^3S_1$ channel in permuted
  ordering.  Some of these orderings can be identified with Wilson or
  Wegner SRG generators when the infrared limit is taken $\lambda
  \rightarrow 0$. We have considered a high-momentum UV
  cutoff $\Lambda=2~{\rm fm}^{-1}$ and different number of grid points
  $N=10,20,30,40,50,100$. The points corresponding to the momentum
  at which the deuteron bound-state eigenvalue is placed on the diagonal
  of the hamiltonian are omitted. We also show the exact phase-shifts
  obtained from the solution of the standard LS equation.}
\label{fig:16}
\end{figure*}

\section{The energy-shift operator}

The most obvious phase-shift definition preserving phase-equivalence
on the grid should involve the spectrum. Fortunately, this was done
long ago by Lifshits and has recently received a lot of attention by
Kukulin {\it el al.} who extended the energy-shift approach to few-nucleon
problems~\cite{kukulin2009discrete,Rubtsova:2010zz}. Their setup
allows to solve scattering problems without ever solving the
scattering equations, since it just involves the energy
eigenvalues. It is important to note that for an $N$-dimensional momentum grid there are $N!$ possible orderings for the eigenvalues of the Hamiltonian obtained from any diagonalization procedure and so the evaluation of phase-shifts using the energy-shift approach necessarily involves a prescription to order the states.

The general result in the presence of $n_B$ bound-states derived by Kukulin {\it el al.} is written as
\begin{eqnarray}
\delta_n^{\rm Kuk} = - \pi \frac{P_{n+n_B}^2-p_n^2}{2 w_n p_n} \;.
\label{deltakuk}
\end{eqnarray}
with $n=1, \cdots, ~N-n_B$. According to this prescription, in order to evaluate the phase-shifts the eigenvalues $P_{n}^2$ obtained from the diagonalization of the Hamiltonian $H$ (arranged in ascending order) must be shifted to the left by $n_B$ positions with respect to the corresponding eigenvalues $p_{n}^2$ of the free hamiltonian $T$. One should note that such a prescription implies that the first $n_B$ eigenvalues (those corresponding to the bound-states) are removed when the shift is implemented. The results obtained by applying Eq. (\ref{deltakuk}) to evaluate the phase-shifts for the toy-model potential in the $^1S_0$ and $^3S_1$ channels with several number of grid points $N$ can be seen in Fig.~(\ref{fig:16}). In the case of the $^1S_0$ channel, which has no bound-state, there is no shift of the eigenvalues $P_{n}^2$ since $n_B=0$ and the prescription works rather well in the entire range of momenta as one can see in the upper-left panel. The situation for the $^3S_1$ channel is different since $n_B=1$ due to the presence of the deuteron bound-state. As we can see in the left-bottom panel, when no shift is applied the low-momenta behavior clearly violates Levinson's theorem. As shown in the upper-right panel, the low-momenta behavior is properly fixed by shifting the eigenvalues according to Kukulin's prescription and looks like fulfilling Levinson's theorem for one bound-state. However, the large momentum behavior is greatly distorted due the mismatch of the free momenta and the eigenvalues generated by the shift. This effect survives in the continuum limit and the upper bending indicates that Levinson's theorem is fulfilled, however, with no bound states. Thus we are faced to the problem of defining an isospectral phase-shift with a proper high-energy behavior.

Clearly, in order to avoid the high-energy mismatch the constant shift implied by Kukulin's formula should not be used. On the other hand, the shifted formula complies to Levinson's theorem at low-energies. Thus, even within the isospectral scenario there seems to be a conflict between high-energy behavior and the fulfillment of Levinson's theorem. Therefore, the question is at what location should the shift of the eigenvalues be applied in order to obtain phase-shifts that have a proper behavior both at low-energies and high-energies.

\section{The SRG induced ordering of states}

As pointed out before, in the case of the SRG evolution with Wilson generator there is only one asymptotically stable final ordering of the eigenvalues, corresponding to the permutation in which the eigenvalues are ordered according to the kinetic energy (i.e., in ascending order). On the other hand, the SRG evolution with Wegner generator allows in principle any asymptotically stable final ordering of the eigenvalues. However, the uniqueness of the solution implies that just one ordering takes place asymptotically for $\lambda\to 0$. In the absence of bound-states, the final ordering for the Wegner generator is the same as for the Wilson generator (ascending order).

We define the SRG-ordered phase-shift for the generator $G$ as follows
\begin{eqnarray}
\delta_n^{G} = - \pi \lim_{\lambda \to 0} \frac{H_{n,n}^{G,\lambda}-p_n^2}{2 w_n p_n} \; .
\end{eqnarray}
If we denote by ${E_n}$ the spectrum of $H_{n,n}^{G,\lambda}$ in ascending order, i.e. $E_{1}  < \cdots < E_{N} $,
we generally have
\begin{eqnarray}
\lim_{\lambda \to 0 }H_{n,m}^{G,\lambda} = \delta_{n,m} E_{\pi(n)} = \delta _{n,m} P^ 2_{\pi(n)} \; ,
\end{eqnarray}
where $\pi(n)$ is one of the $N!$ permutations of the N-plet $(1, \dots,
N)$.

For the Wilson generator, $G_s=T$, one can show that the ascending
order is asymptotically preserved~\cite{Timoteo:2011tt,Arriola:2013gya},
\begin{eqnarray}
\lim_{\lambda \to 0 }H_{n,m}^{{\rm Wil},\lambda} = \delta_{n,m} E_{n} \;,
\label{hwil}
\end{eqnarray}
and thus
\begin{eqnarray}
\delta_n^{\rm Wil} = - \pi \lim_{\lambda \to 0} \frac{H_{n,n}^{{\rm Wil},\lambda}-p_n^2}{2 w_n p_n}= - \pi \frac{P_{n}^2-p_n^2}{2 w_n p_n}
\label{deltawil}
\end{eqnarray}
which corresponds to Kukulin's formula with no shift and thus leads to the violation of Levinson's theorem in the presence of bound-states.

For the Wegner generator case, $G_s=diag(H_s)$,
\begin{eqnarray}
\delta_n^{\rm Weg} = - \pi \lim_{\lambda \to 0} \frac{H_{n,n}^{{\rm Weg},\lambda}-p_n^2}{2 w_n p_n}
\label{eq:ps-weg}
\end{eqnarray}
\begin{figure}[t]
\begin{center}
\includegraphics[width=8.5cm]{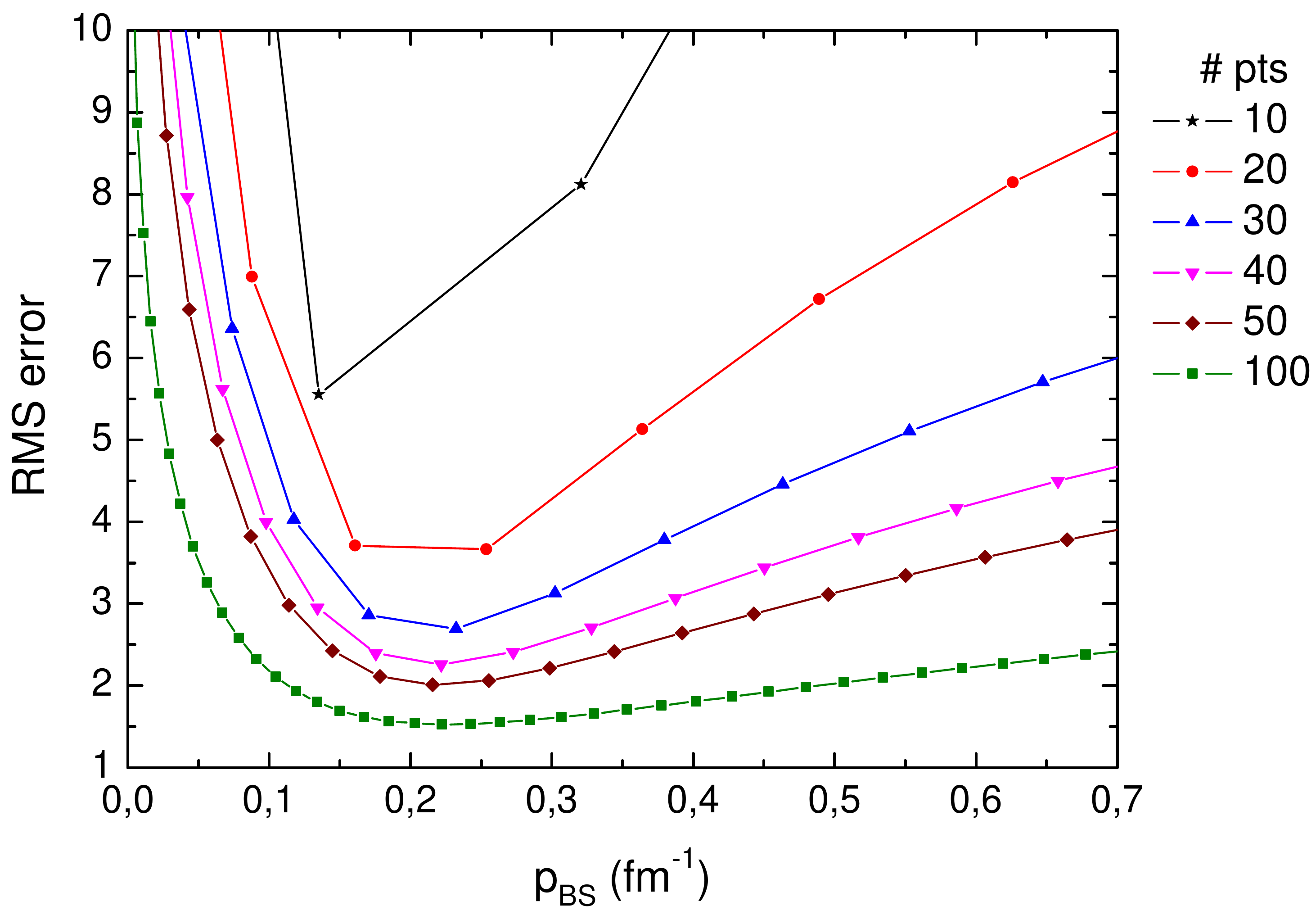}\hspace{0.5cm}
\end{center}
\caption{RMS errors in the phase-shifts (in degrees) for the toy model
  potential in the $^3S_1$ channel evaluated from the energy-shift
  formula by varying the position of the deuteron bound-state
  eigenvalue. We have considered a high-momentum UV cutoff
  $\Lambda=2~{\rm fm}^{-1}$ and different number of grid points
  $N=10,20,30,40,50,100$.}
\label{fig:rms}
\end{figure}

Our analysis of the results obtained for the SRG evolution of the toy-model Hamiltonian, shown in Fig.~(\ref{fig:25}), suggest an alternative prescription to order the eigenvalues when using the energy-shift approach to evaluate the phase-shifts. By placing the bound-state eigenvalue at the position induced by the SRG evolution with Wegner generator in the infrared limit ($\lambda\to 0$), which corresponds to the grid momentum $p_{n_{\rm BS}}$, we have
\begin{eqnarray}
\lim_{\lambda \to 0 }H_{n,m}^{{\rm Weg},\lambda} = \delta_{n,m}
\begin{cases}
P_{n+1}^2 \quad {\rm if} \quad n < n_{\rm BS}  \\
-\gamma^ 2  \, \, \quad {\rm if} \quad n = n_{\rm BS}  \\
P_{n}^2 \quad \quad {\rm if} \quad n < n_{\rm BS}
\end{cases}
\end{eqnarray}
Literal application of this result in Eq.~(\ref{eq:ps-weg}) generates a discontinuity at $\delta_{n_{\rm BS}}$. We can instead just remove the point at the position $n=n_{\rm BS}$ corresponding to the location of the bound-state eigenvalue, similar to what is done in Kukulin's prescription, or interpolate between the neighboring values, taking $P_{n_{\rm BS}}^2 \to \bar P_{n_{\rm BS}}^2=(P_{n_{\rm BS}+1}^2+P_{n_{\rm BS}-1}^2)/2$. This yields
\begin{eqnarray}
\bar \delta_n^{\rm Weg} =
\begin{cases}
- \pi \frac{P_{n+1}^2-p_n^2}{2 w_n p_n} \quad  {\rm if} \quad n < n_{\rm BS}  \\
- \pi \frac{\bar P_n^2-p_n^2}{2 w_n p_n} \qquad {\rm if} \qquad n=n_{\rm BS} \\
- \pi \frac{P_n^2-p_n^2}{2 w_n p_n} \qquad {\rm if} \quad n > n_{\rm BS}
\end{cases}
\label{deltaweg}
\end{eqnarray}

\begin{table}[ht]
\caption{Comparison between the position of the deuteron bound-state which minimizes the RMS errors in the phase-shifts, $p_{n_{\rm BS}}^{\rm opt}$, and the position induced by the SRG evolution with Wegner generator in the infrared limit, $p_{n_{\rm BS}}^{\rm weg}$, for different number of grid points $N=10,20,30,40,50,100$.}
\vskip 0.5cm
\centering
\begin{tabular}{c c c}
\hline\hline
\\
N & $p_{n_{\rm BS}}^{\rm opt}$ (${\rm fm}^{-1}$) & $p_{n_{\rm BS}}^{\rm weg}$(${\rm fm}^{-1}$) \\ \\
\hline\hline \\
10 & 0.135 & 0.321  \\
20 & 0.254 & 0.254 \\
30 & 0.232 & 0.170  \\
40 & 0.221 & 0.175  \\
50 & 0.215 & 0.145  \\
100& 0.222 & 0.104  \\ \\
\hline
\end{tabular}
\label{table:rmsweg}
\end{table}

In this way, we get an ordering prescription in which {\it only} the eigenvalues corresponding to momenta $p_n < p_{n_{\rm BS}}$ are shifted one position to the left, unlike Kukulin's prescription in which {\it all} eigenvalues are shifted. As pointed before, the position of the bound-state eigenvalue induced by the SRG evolution with Wegner generator changes when using different values for the number of grid points $N$ and so the momentum $p_{n_{\rm BS}}$ below which the shift is applied. In the bottom-right panel of Fig.~(\ref{fig:16}) we show the phase-shifts evaluated from Eq.~(\ref{deltaweg}) for different number of grid points $N$, compared to the exact results obtained from the solution of the standard LS equation. As one can see, both low-energy and high-energy behaviors are correct within the expected uncertainties of the finite grid. The good job performed by the SRG evolution with Wegner generator in properly locating the momentum $p_{n_{\rm BS}}$ at which the bound-state eigenvalue must be placed when using the energy-shift approach can be traced to the decoupling of the bound-state from the low-momentum scales in the infrared limit.

Thus, we find that remarkably the ordering of the eigenvalues induced by the SRG evolution with Wegner generator in the infrared limit provides a prescription which allows to obtain isospectral phase-shifts that fulfill Levinson's theorem at low-momenta and have a proper behavior at high-momenta. However, such an ordering does not correspond in general to the optimal one. We have evaluated the phase-shifts for the toy model potential in the $^3S_1$ channel from the energy-shift formula by varying the position of the deuteron bound-state eigenvalue $p_{n_{\rm BS}}$ and compared to the exact results obtained from the solution of the standard LS equation. In Fig.~(\ref{fig:rms}) we show the plots corresponding to the RMS errors {\it versus} $p_{n_{\rm BS}}$ computed for several number of grid points $N$. As one can see in Table~\ref{table:rmsweg}, the position of the deuteron bound-state which minimizes the RMS errors, $p_{n_{\rm BS}}^{\rm opt}$, is different from the position induced by the SRG evolution with Wegner generator in the infrared limit, $p_{n_{\rm BS}}^{\rm weg}$. Since the SRG evolution with distinct generators induces different isospectral flows in the presence of bound-states, it is plausible to conceive that a specific generator may be found which leads to the optimal ordering. It is also interesting to note that in the continuum limit the position of the deuteron bound-state which minimizes the RMS errors seems to approach the characteristic deuteron momentum scale $\gamma = 0.23~{\rm fm}^{-1}$. Of course, it must be verified through explicit calculations if this a general result, which holds for any weakly or strongly coupled bound-state.

\section{Conclusions}

We have unveiled a remarkable connection between the SRG evolution for
a generic generator and the Levinson's theorem on a finite momentum
grid, where the scattering problem turns into a bound state
problem. So some naive relations such as the phase-equivalence of the
transformation depend on the very definition of the phase-shift and
certainly do not hold for the customary Lippmann-Schwinger
definition. An isospectral definition is based on an energy shift due
to the interaction and is phase-equivalent along the SRG trajectory,
but different generators provide different eigenvalue orderings and
fulfilling Levinson's theorem depends on knowledge of the location of
a bound state scale in momentum space.  We have seen that while the
Wilson generator induces an ordering contradicting Levinson's theorem,
the Wegner generator does a much better job, but still underestimates
the relevant momentum scale. The main handicap to the general analysis
seems to be that within the matrix formulation it is difficult to
profit of the specific information embodied in quantum mechanical
Hamiltonians.  A more rigorous discussion will probably implement
asymptotic and analytic features of the Hamiltonian as a function of
the momentum and in the complex plane. We remind that the standard
proof of Levinson's theorem in the continuum makes extensive use of
these features~\cite{Ma:2006zzc}.

For the case under study we restrict ourselves to the one single bound
state situation, which actually corresponds to the case of interest in
the two nucleon problem. We have also checked that our prescription
works also for realistic potentials. Our results should find a sensible
generalization for more bound states, and we leave this study for
future investigation.

%Of course, we have used a simple discretization approach
%which has not been optimized for a finite number of grid points, and
%it would also be interesting to improve on this side.

%While our conjecture works rather well and
%highlights specific virtues of the Wegner generator in the infrared
%limit we have no proof that when there are no bound states the
%ordering for the Wegner remains the same as in Wilson case.

\section*{Acknowledgements}

E.R.A. was supported by Spanish DGI (grant FIS2011-24149) and Junta de
Andaluc\1a (grant FQM225). S.S.  was supported by FAPESP and V.S.T.
by FAEPEX/PRP/UNICAMP, FAPESP and CNPq. Computational power provided
by FAPESP grant 2011/18211-2.

\bibliographystyle{elsarticle-num}
%\bibliography{levinson}

\begin{thebibliography}{10}
\expandafter\ifx\csname url\endcsname\relax
  \def\url#1{\texttt{#1}}\fi
\expandafter\ifx\csname urlprefix\endcsname\relax\def\urlprefix{URL }\fi
\expandafter\ifx\csname href\endcsname\relax
  \def\href#1#2{#2} \def\path#1{#1}\fi

\bibitem{Bogner:2006pc}
S.~Bogner, R.~Furnstahl, R.~Perry, {Similarity Renormalization Group for
  Nucleon-Nucleon Interactions}, Phys.Rev. C75  061001.

\bibitem{Furnstahl:2013oba}
R.~Furnstahl, K.~Hebeler, {New applications of renormalization group methods in
  nuclear physics}, Rept.Prog.Phys. 76 (2013) 126301.
\newblock \href {http://arxiv.org/abs/1305.3800} {\path{arXiv:1305.3800}},
  \href {http://dx.doi.org/10.1088/0034-4885/76/12/126301}
  {\path{doi:10.1088/0034-4885/76/12/126301}}.

\bibitem{Timoteo:2011tt}
V.~Timoteo, S.~Szpigel, E.~Ruiz~Arriola, {Symmetries of the Similarity
  Renormalization Group for Nuclear Forces}, Phys.Rev. C86 (2012) 034002.
\newblock \href {http://arxiv.org/abs/1108.1162} {\path{arXiv:1108.1162}},
  \href {http://dx.doi.org/10.1103/PhysRevC.86.034002}
  {\path{doi:10.1103/PhysRevC.86.034002}}.

\bibitem{Arriola:2013nja}
E.~Ruiz~Arriola, V.~Timoteo, S.~Szpigel, {Nuclear Symmetries of the similarity
  renormalization group for nuclear forces}, PoS CD12 (2013) 106.
\newblock \href {http://arxiv.org/abs/1302.3978} {\path{arXiv:1302.3978}}.

\bibitem{Dainton:2013axa}
B.~Dainton, R.~Furnstahl, R.~Perry, {Universality in Similarity Renormalization
  Group Evolved Potential Matrix Elements and T-Matrix Equivalence}, Phys.Rev.
  C89 (2014) 014001.
\newblock \href {http://arxiv.org/abs/1310.6690} {\path{arXiv:1310.6690}},
  \href {http://dx.doi.org/10.1103/PhysRevC.89.014001}
  {\path{doi:10.1103/PhysRevC.89.014001}}.

\bibitem{Perez:2013jpa}
R.~N. Perez, J.~Amaro, E.~R. Arriola, {Coarse grained Potential analysis of
  neutron-proton and proton-proton scattering below pion production threshold},
  Phys.Rev. C88 (2013) 064002.
\newblock \href {http://arxiv.org/abs/1310.2536} {\path{arXiv:1310.2536}},
  \href {http://dx.doi.org/10.1103/PhysRevC.88.064002}
  {\path{doi:10.1103/PhysRevC.88.064002}}.

\bibitem{Bulgac:2013mz}
A.~Bulgac, M.~M. Forbes, {On the use of the Discrete Variable Representation
  Basis in Nuclear Physics}, Phys.Rev. C87 (2013) 051301.
\newblock \href {http://arxiv.org/abs/1301.7354} {\path{arXiv:1301.7354}},
  \href {http://dx.doi.org/10.1103/PhysRevC.87.051301}
  {\path{doi:10.1103/PhysRevC.87.051301}}.

\bibitem{Arriola:2013gya}
E.~Ruiz~Arriola, S.~Szpigel, V.~Timoteo, {Fixed points of the Similarity
  Renormalization Group and the Nuclear Many-Body Problem}, Few Body Syst.\href
  {http://arxiv.org/abs/1310.8246} {\path{arXiv:1310.8246}}, \href
  {http://dx.doi.org/10.1007/s00601-014-0858-7}
  {\path{doi:10.1007/s00601-014-0858-7}}.

\bibitem{muga1989stationary}
J.~Muga, R.~Levine, Stationary scattering theories, Physica Scripta 40~(2)
  (1989) 129.

\bibitem{kukulin2009discrete}
V.~I. Kukulin, V.~N. Pomerantsev, O.~A. Rubtsova, Discrete representation of
  the spectral shift function and the multichannel s-matrix, JETP letters
  90~(5) (2009) 402--406.

\bibitem{Rubtsova:2010zz}
O.~Rubtsova, V.~Kukulin, V.~Pomerantsev, A.~Faessler, {New approach toward a
  direct evaluation of the multichannel multienergy S matrix without solving
  the scattering equations}, Phys.Rev. C81 (2010) 064003.
\newblock \href {http://dx.doi.org/10.1103/PhysRevC.81.064003}
  {\path{doi:10.1103/PhysRevC.81.064003}}.

\bibitem{Fukuda:1956zz}
N.~Fukuda, R.~Newton, {Energy Level Shifts in a Large Enclosure}, Phys.Rev. 103
  (1956) 1558--1564.
\newblock \href {http://dx.doi.org/10.1103/PhysRev.103.1558}
  {\path{doi:10.1103/PhysRev.103.1558}}.

\bibitem{DeWitt:1956be}
B.~S. DeWitt, {Transition from discrete to continuous spectra}, Phys.Rev. 103
  (1956) 1565--1571.
\newblock \href {http://dx.doi.org/10.1103/PhysRev.103.1565}
  {\path{doi:10.1103/PhysRev.103.1565}}.

\bibitem{Luscher:1985dn}
M.~Luscher, {Volume Dependence of the Energy Spectrum in Massive Quantum Field
  Theories. 1. Stable Particle States}, Commun.Math.Phys. 104 (1986) 177.
\newblock \href {http://dx.doi.org/10.1007/BF01211589}
  {\path{doi:10.1007/BF01211589}}.

\bibitem{Luscher:1990ux}
M.~Luscher, {Two particle states on a torus and their relation to the
  scattering matrix}, Nucl.Phys. B354 (1991) 531--578.
\newblock \href {http://dx.doi.org/10.1016/0550-3213(91)90366-6}
  {\path{doi:10.1016/0550-3213(91)90366-6}}.

\bibitem{Ma:2006zzc}
Z.-Q. Ma, {The Levinson theorem}, J.Phys. A39 (2006) R625--R659.
\newblock \href {http://dx.doi.org/10.1088/0305-4470/39/48/R01}
  {\path{doi:10.1088/0305-4470/39/48/R01}}.

\bibitem{Kehrein:2006ti}
S.~Kehrein, {The flow equation approach to many-particle systems}, Springer,
  2006.

\bibitem{wegner1994flow}
F.~Wegner, {Flow-equations for Hamiltonians}, Annalen der physik 506~(2) (1994)
  77--91.

\bibitem{Glazek:1994qc}
S.~D. Glazek, K.~G. Wilson, {Perturbative renormalization group for
  Hamiltonians}, Phys. Rev. D49 (1994) 4214--4218.

\bibitem{Szpigel:2010bj}
S.~Szpigel, V.~S. Timoteo, F.~d.~O. Duraes, {Similarity Renormalization Group
  Evolution of Chiral Effective Nucleon-Nucleon Potentials in the Subtracted
  Kernel Method Approach}, Annals Phys. 326 (2011) 364--405.
\newblock \href {http://arxiv.org/abs/1003.4663} {\path{arXiv:1003.4663}},
  \href {http://dx.doi.org/10.1016/j.aop.2010.11.007}
  {\path{doi:10.1016/j.aop.2010.11.007}}.

\bibitem{Arriola:2013era}
E.~Ruiz~Arriola, S.~Szpigel, V.~Timoteo, {Implicit vs Explicit Renormalization
  and Effective Interactions}, Phys.Lett. B728 (2014) 596--601.
\newblock \href {http://arxiv.org/abs/1307.1231} {\path{arXiv:1307.1231}},
  \href {http://dx.doi.org/10.1016/j.physletb.2013.12.038}
  {\path{doi:10.1016/j.physletb.2013.12.038}}.

\bibitem{Arriola:2013yca}
E.~R. Arriola, S.~Szpigel, V.~S. Timoteo, {Implicit vs Explicit renormalization
  of the $NN$ force}, Few Body Syst.\href {http://arxiv.org/abs/1310.8526}
  {\path{arXiv:1310.8526}}, \href {http://dx.doi.org/10.1007/s00601-014-0811-9}
  {\path{doi:10.1007/s00601-014-0811-9}}.

\bibitem{Glazek:2008pg}
S.~D. Glazek, R.~J. Perry, {The impact of bound states on similarity
  renormalization group transformations}, Phys.Rev. D78 (2008) 045011.
\newblock \href {http://arxiv.org/abs/0803.2911} {\path{arXiv:0803.2911}},
  \href {http://dx.doi.org/10.1103/PhysRevD.78.045011}
  {\path{doi:10.1103/PhysRevD.78.045011}}.

\bibitem{Wendt:2011qj}
K.~Wendt, R.~Furnstahl, R.~Perry, {Decoupling of Spurious Deep Bound States
  with the Similarity Renormalization Group}, Phys.Rev. C83 (2011) 034005.
\newblock \href {http://arxiv.org/abs/1101.2690} {\path{arXiv:1101.2690}},
  \href {http://dx.doi.org/10.1103/PhysRevC.83.034005}
  {\path{doi:10.1103/PhysRevC.83.034005}}.

\end{thebibliography}

\end{document}